# TECNOLOGIA MÓVEL: UMA TENDÊNCIA, UMA REALIDADE.

Carlos Augusto Almeida Alcantara[1]
Anderson Luiz Nogueira Vieira[2]

**Resumo**

Atualmente a mobilidade (*Mobile*), apresenta-se como uma das grandes inovações históricas da revolução tecnológica. A partir da primeira década do século XXI, presencia-se um impacto profundo e fundamental em boa parte dos setores econômicos, mas nada comparado ao que ocorreu no setor de Tecnologia da Informação (TI) que, vem acrescentando ao mercado, uma gama de novidades relativas à infraestrutura da computação móvel como hardwares, softwares, redes de computadores, etc. A partir de um equipamento de mão, como um dispositivo celular, iPad, Smartphone, é possível conectar-se ao mundo digital, pessoas, clientes, parceiros de negócios, etc. Tendo em vista a infinita gama de informações, serviços e recursos disponíveis no mundo eletrônico, considera-se que poucos são aqueles, mercados e pessoas, que querem ficar para trás. O interesse neste canal de comunicação se transforma em uma nova estratégia não só de comercialização como de comunicação.

Os dispositivos móveis estão cada vez mais sofisticados e permite o acesso a WEB. Pensando nesse contexto, é que os fabricantes de componentes eletrônicos se enfrentam numa guerra tecnológica pela disputa de um ambiente competitivo que se abre para quem colocar no mercado, o produto mais atraente, interativo e versátil.

**Palavras Chaves:** Mobile, tecnologia, smartphone, mobilidade, comunicação, redes.

[1]
Graduando em Redes de Computadores pela Faculdade Estácio de Sá de Juiz de Fora – MG;
E-mail: gutoalcantara@gmail.com

[2]
Pós-graduado em Ciência da Computação pela Universidade Federal de Viçosa.
Pós-graduado em Redes de Computadores pela Universidade Estácio de Sá
Graduado em Tecnologia em Processamento de Dados pelo CES/JF
E-mail: anderson.vieira@gmail.com

**Abstract**

Currently, mobility presents itself as a major innovation in historic technological revolution. From the first decade of this century, nothing compares to what happened in the field of Information Technology (IT), which is adding to market a range of news relating to infrastructure such as mobile computing hardware, software, computer networks, etc.. From a handheld as a wireless device, iPad, Smartphone, you can connect to the digital world, people, customers, business partners, etc.. Given the infinite range of information, services and resources available in the electronic world, it is considered that few are those, markets and people who want to be left behind. The interest in this channel of communication becomes not only a new strategy of marketing and communications.

Mobile devices are becoming more sophisticated and allows access to the web. Thinking in this context is that manufacturers of electronic components face each other in a war over the disputed technology to a competitive environment that is open to those who put on the market the product more attractive, interactive and versatile.

Keywords: Mobile, technology, smartphone, mobility, communication, networking.

**Introdução**

Podemos definir *Tecnologia Móvel* como a forma de acessar a internet e outros recursos computacionais por meio de dispositivos móveis, tais como, celulares, iPhone, iPod, iPad, notebooks, smartpads, dentre outros. A cada dia, um número maior de pessoas interessa-se pela mobilidade, o fácil acesso às informações em qualquer lugar, com alcance amplo a qualquer hora, se conectando de forma fácil e rápida a outros dispositivos móveis, localizando pessoas, produtos e serviços personalizados. Estes são os fatores que impulsionam a internet móvel a se estruturar e crescer rapidamente para adaptar às modernidades e necessidades dos usuários finais, bem como das organizações.

Mediante a realidade demandada por serviços e recursos móveis, começam a surgir no mercado sites desenvolvidos para promover a liberdade no uso e consumo de informações por parte dos usuários finais que adquirem aparelhos como os smartphones que possuem facilidade de uso e recursos cada vez mais modernos. Muitos sites e soluções para internet atualmente já tem como foco sua usabilidade em dispositivos móveis. Assim o uso da internet através dos Smartphones difere do uso no computador sob vários aspectos. A diferença, na verdade é uma das vantagens, consiste em não necessitar de um computador com teclado e mouse conectados fisicamente. Por meio do uso dos dispositivos móveis, a liberdade de locomoção com acesso a internet é o ponto favorável e que desperta o interesse dos usuários finais e atração comercial de campanhas publicitárias. A mobilidade assim permite ter em mãos serviços, informações, comunicação e entretenimento. No caso dos serviços incluem as consultas bancárias, previsões do tempo, notícias, redes sociais, operações em tempo real. Todo esse conteúdo é obtido por meio de pesquisas, a endereços, telefones, promoções e produtos. Para a comunicação a possibilidade de comunicar, interagir e compartilhar, faz com que pessoas se encontrem, troquem ideias, realizem negócios e cooperem uns com os outros.

Neste contexto, este artigo procura trazer o entendimento de tudo aquilo que se passa em nossa volta de forma a ser absorvido com mais rapidez e clareza. Para tanto está disposto da seguinte forma: o Capitulo 2 faz um comparativo entre os serviços convencionais e móveis; o Capítulo 3 fala sobre o crescimento da tecnologia móvel; O capítulo 4 mostra a concretização da tecnologia móvel nas grandes corporações; o capítulo 5 evidencia o crescimento no comércio eletrônico; o capitulo 6 fala sobre a Banda Larga Móvel no Brasil, sua utilização e seu crescimento. Por fim, têm-se as conclusões finais.

## 2. INTERNET CONVENCIONAL VERSUS TECNOLOGIA MÓVEL

Ambas as tecnologias apresentam limitações e possibilidades, que se adaptam e moldam as oportunidades e os avanços. Segundo (Pearson Education, 2004), a infraestrutura da computação móvel no que diz respeito a Hardware é composta por: Telefones celulares, Teclado acoplável, PDAs, Pagers interativos, Notebooks, iPhone, iPod, Smartpads.

> Brasil, 2009. Ação de mobile marketing utilizando mídia off-line. Um anúncio de página inteira veiculado em um jornal impresso convidava os leitores a enviar um SMS gratuito para um LA (Large Account – número para envio de SMSs) para obter um desconto. O SMS de retorno continha um click-to-call para o 0800 do televendas e um cupom que poderia ser inserido no hotsite web, que também era divulgado no SMS. (Praestro Convergence - ebook Mobile Marketing)

As Figuras 1, 2 e 3, mostram exemplos de sites para uso no celular (esquerda) e desktop (direita).

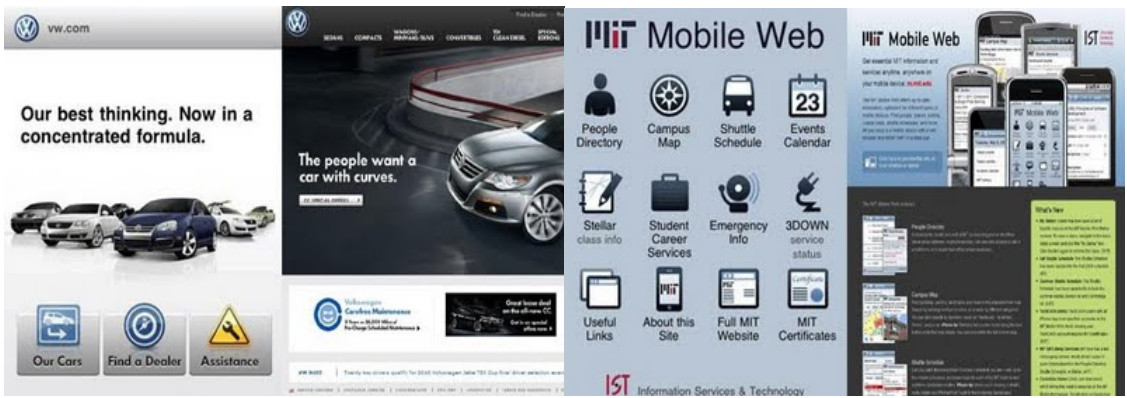

Figura 1 – Exemplos de sites para Celular e Desktop

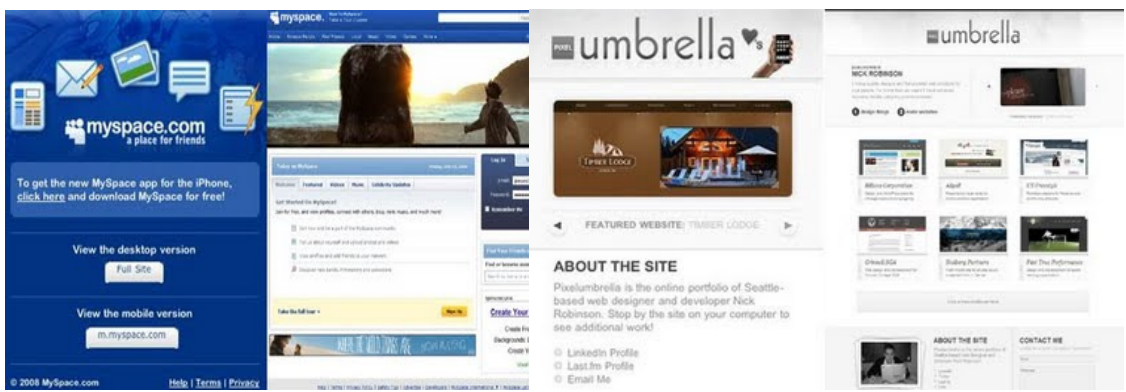

Figura 2 – sites no formato celular e desktop.

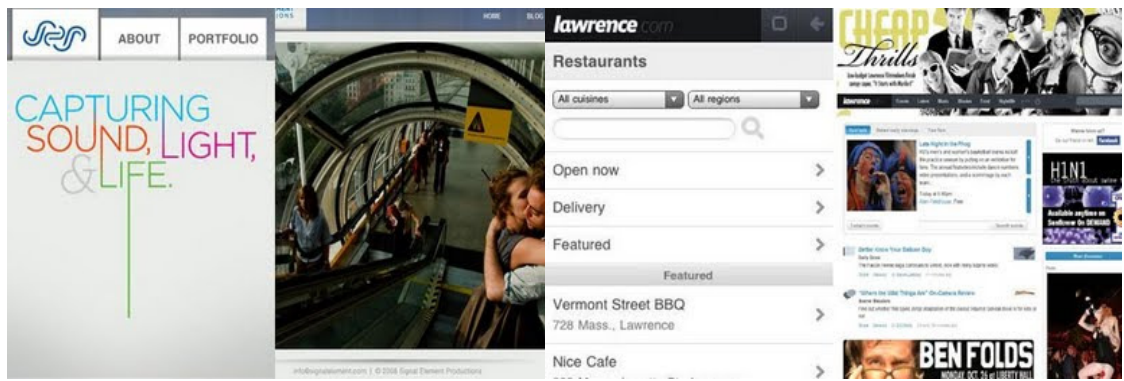

Figura 3 – sites para celular e desktop.

## 3. TECNOLOGIA MÓVEL E TODO SEU CRESCIMENTO

Hoje em dia o usuário tem a vantagem não só de acompanhar o crescimento tecnológico como de usufruir todas as suas vantagens. A cada momento, o mercado se vê na obrigação de melhorar seus atendimentos e vantagens para conseguir uma melhor sedução do seu público-alvo e até mesmo, de seu futuro mercado. Com isso precisa incrementar em seu meio de trabalho um jeito de chamar a atenção e até mesmo de mostrar ao cliente que em momento algum, ele deve se preocupar em levar dinheiro para fazer uma simples compra, de se apavorar por não saber onde está de saber se tem algum dinheiro em sua conta para que possa efetuar uma transação comercial ou até mesmo, de sair de casa para fazer o pagamento de uma conta. Com o crescimento tecnológico do mercado móvel, o empresário está proporcionando tudo isso ao seu cliente. Atualmente, com o crescimento no uso de tecnologias móveis, a facilidade de uma pessoa obter um celular (Smartphone) de qualidade alguns serviços podem ser implementados de forma bastante abrangente.

> Os mercados de telefonia móvel e internet sofreram intensa agitação no último ano. De acordo com dados da Agência Nacional de Telecomunicações, o país chegou a recordes 147 milhões de telefones celulares ativos. Além disso, 2008 também será lembrado como o ano em que a portabilidade chegou ao Brasil e a disputa para fidelizar os clientes aumentou consideravelmente.
> Diante desse cenário e do crescimento da publicidade cada vez mais focada em nichos sociais, a expansão do Mobile Marketing é uma das principais tendências para 2009. De acordo com o gerente de marketing da Wapja.net, Guilherme Lara, mobile marketing "é a mídia mais íntima do usuário e que está 24h ao seu lado. ( LARA, Guilherme, Mobile Marketing terá crescimento vertiginoso em 2009, 2009)

## 3.1 SERVIÇOS DE LOCALIZAÇÃO

Algumas empresas atualmente facilitam a vida de pessoas que por diversas razões viajam muito, com serviços localização em sistema GPS ou Global Positioning System (Sistema de Posicionamento Global), indicando em tempo real a sua exata localização, além de mostrar também informações de pontos históricos e locais turísticos da região. Com um simples apertar de teclas no Smartphone ou celular, o cliente tem acesso a um navegador completo e detalhado com ruas, estradas pavimentadas, caminhos e trilhas, pontos de fiscalização e até mesmo acesso a hospitais e restaurantes, conforme apresentado na Figura 4.

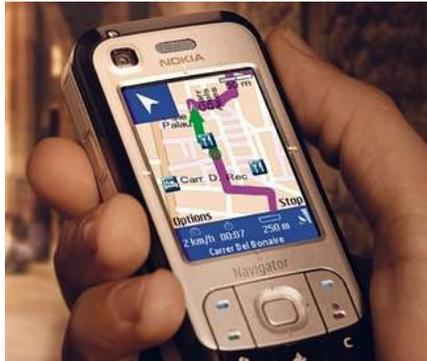
Figura 4 - telefone NOKIA 6110 com GPS

Mas, o serviço de localização não se resume apenas na localização pessoal em relação ao ambiente, a tecnologia está em uma fase tão avançada que já é possível localizar pessoas bastando que elas portem um celular no bolso. Hoje já se pode ter certeza de que sua filha está mesmo na casa de sua amiguinha de escola ou se o seu amigo está realmente preso no trânsito. O cliente receberá uma mensagem de tempo em tempo perguntando se ele deseja ser localizado durante certo período de tempo. Caso o cliente não responda por um tempo, o sistema automaticamente é colocado como o cliente querendo ser localizado, no caso de sequestros e acidentes impossibilitando-o deter qualquer acesso ao celular.

O cliente também poderá encontrar lojas de determinado ramo de atividade em qualquer lugar onde esteja em questão de segundos, e essa loja, sendo credenciada com sistema de vendas por celular, pode até efetuar sua compra sem ter de sair do local onde está.

## 3.2 COMPRAS

A atual situação tecnológica é a de que não é preciso mais sair de casa para quase nada, muito menos para fazer compras. Como algumas pessoas não têm tempo para ir às compras rotineiras e sempre precisam parar seus afazeres para poder fazê-las, as empresas estão disponibilizando de sistemas de venda por telefonia celular. Agora o cliente se cadastra em uma empresa onde aceita esse tipo de venda e basta ligar para a empresa e fazer o pedido desejado. Logo em seguida chega a seu celular uma mensagem com o código do estabelecimento e código da compra, onde o cliente manda uma mensagem para a empresa prestadora do serviço de venda e fornece os códigos recebidos da empresa onde efetuou a compra e, pronto, compra efetuada com êxito. Em são Paulo, há ainda padarias e restaurantes onde o cliente entra no estabelecimento e recebe um cartão com um número, que corresponde à sua comanda. Ao finalizar sua compra é só mandar uma mensagem com o número correspondente ao cartão e seguir as instruções no próprio celular. Nesse caso, o cliente em momento algum teve de se preocupar com fila para pagar ou mesmo ter a certeza se seu cartão de crédito iria passar.

O comércio móvel vem ajudando o setor financeiro, onde empresários fazem compras de ações através de dispositivos móveis. Para isso, basta instalar um aplicativo no celular e ter acesso à cotação diária das ações, como se pode ver na Figura 5.

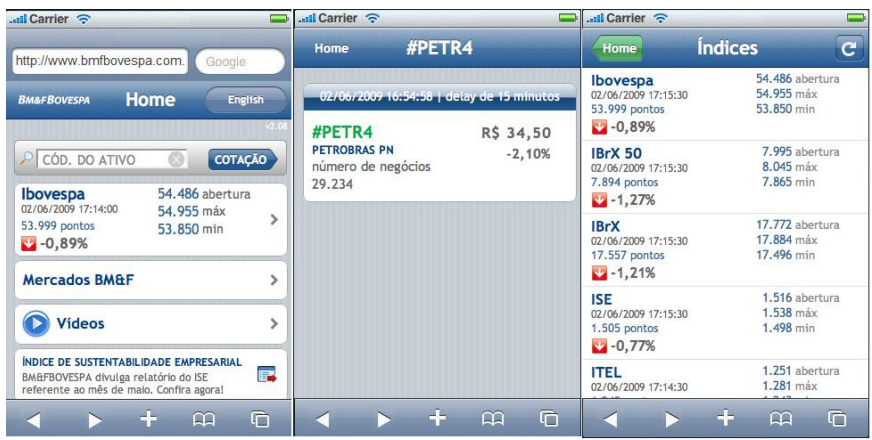

Figura 5: Telas do Bovespa Móvel

Outro serviço muito utilizado é a compra de passagens aéreas podendo até mesmo fazer *check-in* através do celular. O cliente pode ver os horários, paradas, serviços e alternativas. Até o momento, empresas estão apenas cedendo esse serviço para clientes de ponte aérea, mas a intenção é chegar a ampliar esse serviço para todos por conta do atual crescimento da tecnologia. O cliente só tem de entrar 24 horas antes do voo para poder acessar esse serviço, conforme demonstrado na Figura 6.

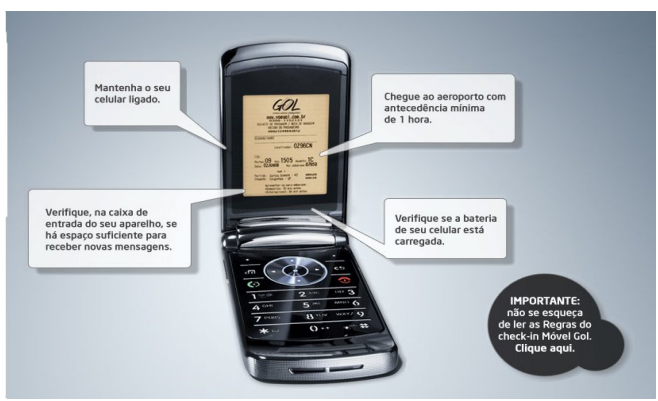

Figura 6: Demonstração de *Check-in Mobil.*

**3.3 COMODIDADE**

O dispositivo móvel se tornou uma ferramenta tão eficiente que deixou de ser utilizado apenas para ligações ou envio de SMS. Segundo LEMOS (2004, p. 6) "O celular passa a ser um 'teletudo', um equipamento que é ao mesmo tempo telefone, máquina fotográfica, televisão, cinema, receptor de informações jornalísticas, difusor de e-mails e SMS7, WAP8, atualizador de sites (moblogs), localizados por GPS, tocador de música (MP3 e outros formatos), carteira eletrônica, fazendo com que a utilização de Notebooks e Netbooks que até pouco tempo eram utilizados como dispositivos móveis. A maior parte dos bancos, hoje em dia, utiliza sistema Bank Phone, que ajuda o cliente a resolver qualquer problema com apenas um telefonema, esteja ele onde estiver. A partir disso, não precisa em momento algum enfrentar filas para se ver livre de contas e problemas financeiros a serem resolvidos.

Com o Smartphone o cliente não precisa mais de um computador para poder trabalhar, basta se conectar a internet e fazer tudo que faria em um escritório só que de qualquer lugar do mundo. A implantação de novas tecnologias na área de telefonia está tornando mais fácil resolver problemas de trabalho, pessoais e até monitorar seguranças de casas e empresas. Agora o usuário passa a ser seu próprio escritório 24 horas por dia e sem gastos adicionais. As pessoas podem fazer videoconferências, pois, uma grande maioria de modelos dispõe de duas câmeras e ainda podem ser usados como webcam.

*Para que comprar jornal se temos um Smartphone?* A partir de aplicativos instalados em um celular, o cliente pode acessar milhões de notícias diariamente de diversos jornais e sites de notícias. A utilização dependerá apenas de um ambiente onde tenha cobertura de sua operadora e sem se preocupar com tempo o próprio aplicativo já se atualiza de tempo em tempo deixando-o informado de qualquer fato no mundo inteiro. É o caso do jornal esportivo Lance, que é atualizado de hora em hora em uma interface completamente fácil de utilizar, como visto na Figura 7.

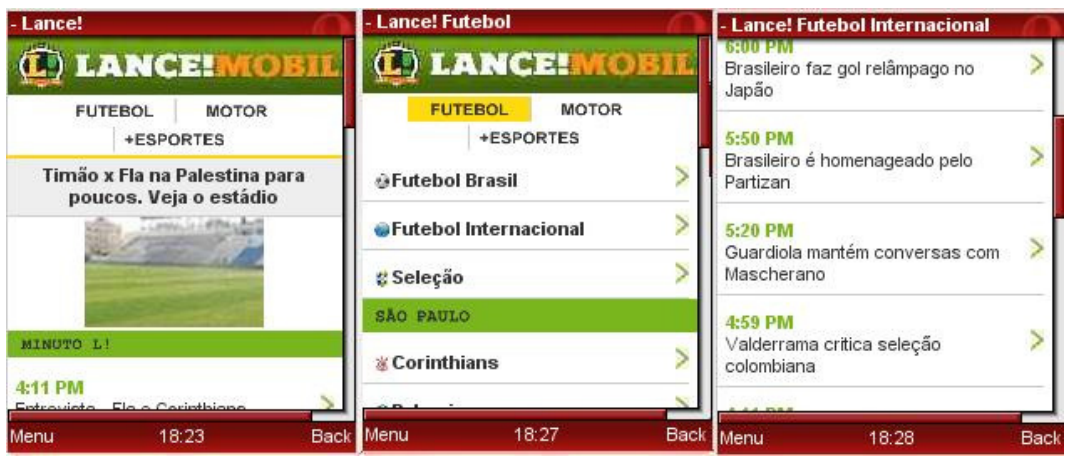

Figura 7: Telas do jornal Lance! para celular.

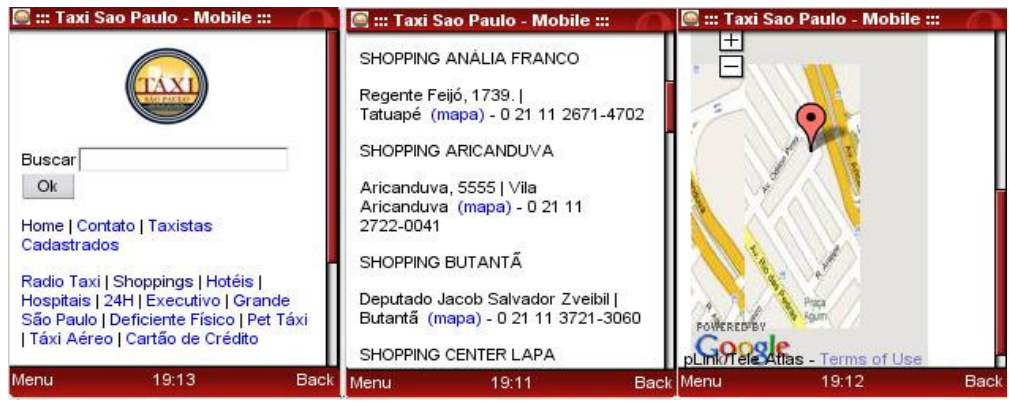

Figura 8: Telas do "Taxi São Paulo - Mobile"

O serviço móvel nos auxilia tanto que temos até serviços de táxi via internet móvel. Em São Paulo, já se encontra em funcionamento pleno um serviço de *Taxi-Mobile*, onde tanto o cliente quanto o próprio taxista pode utilizar-se dele. No caso do cliente, pode usá-lo para chamar um táxi, para encontrar um determinado local na cidade ou uma localidade. E no caso do taxista pode utilizá-lo para encontrar pessoas que estão esperando em algum lugar da cidade ou até mesmo para se localizar e chegar a um ponto desejado. Nesse serviço também dá a possibilidade de o cliente conseguir endereços de milhares de empresas, shoppings, restaurantes, hotéis e tudo mais, como mostra a Figura 8.

> Está mais do que provado que tanto o celular quanto o SMS já fazem parte da vida das pessoas. Exemplos? Um jogador de futebol francês foi cortado da Copa do Mundo porque paquerou a mulher do técnico via SMS. Um piloto de Fórmula 1 foi demitido por ter enviado de forma errada uma mensagem de texto. Pode parecer incomum, mas já existem pessoas contratando profissionais por SMS. Até o presidente dos EUA, Barack Obama, anunciou o seu candidato a vice via SMS. (CASTELO, Marcelo, 2009).

### 3.4 SEGURANÇA

A questão que ainda incomoda muita gente é saber se esses serviços são realmente seguros. De acordo com muitos serviços, é um serviço completamente seguro sim, mas como todo equipamento informatizado terá sempre o problema de vírus. Algumas empresas estão se equipando e fazendo parcerias ao ponto de não deixar "margem para o azar".

Algumas empresas estão com sistemas onde o cliente cadastra o telefone completo e só pode fazer os pagamentos naquele aparelho, caso alguém descubra a senha não terá como fazer nada. Outras empresas estão fazendo além de cadastrar os aparelhos ainda pedem a senha e login todas as vezes que o usuário tentar utilizar do serviço, pois no caso de perda ou roubo do celular, ainda terão de saber os dados para que possa ter acesso aos serviços.

## 4. CONCRETIZANDO A TECNOLOGIA MÓVEL

### 4.1. AMAZON.COM Permite compras pelo celular

Usuários de celular já podem fazer compras na Amazon.com nos Estados Unidos por meio do aparelho. A empresa lançou um sistema que permite o acesso a uma versão do seu site pelo telefone. Os consumidores devem cadastrar previamente seus dados, incluindo endereço de entrega e acessar o serviço usando o cartão de crédito. A novidade lançada por uma das líderes do varejo on-line é mais uma demonstração da posição que as transações por celular ocupará no mercado.

### 4.2. Coca-Cola autoriza Venda de Bebidas pelo Celular

A Coca-Cola anuncia uma novidade móvel para o mercado de bebidas. A partir de agora, será possível comprar e pagar refrigerantes, sucos, chás e energéticos em máquinas da marca, denominadas vending machines, utilizando o celular. Quem traz a novidade para o Brasil é a Refrigerantes Minas Gerais (Remil), fabricante dos produtos Coca-Cola no Estado, em parceria com a Coca-Cola Company, M-Pay (Mobile Payment System), Telemig Celular e Visa.

### 4.3. Consumidor poderá comprar passagens aéreas pelo Celular

14 de maio de 2007 às 07:00 - InfoMoney

Brasileiros ganharam um estímulo para fazer compras pelo celular. Após a consultoria TNS Interscience ter divulgado pesquisa na qual se afirmava que nenhum consumidor do País tinha esse costume, a operadora Tim e a companhia aérea Gol lançaram uma parceria de aquisições de passagens por meio de aparelho móvel.

O levantamento da TNS mostrou que 63% da população sabem que já existem tecnologias do tipo.

O cliente Tim/Gol que quiser utilizar a nova facilidade deve acessar o Portal Tim Wap, comprar o bilhete do destino escolhido e ficar atento ao código que será gerado após a operação. É esse dado que deverá ser entregue no balcão da companhia. Além disso, assim como ocorre com a internet, será possível realizar o check-in por meio do celular.

Por meio de nota, as empresas garantiram que serão mantidas tarifas e condições de pagamento para os clientes da nova tecnologia. O serviço está disponível para qualquer usuário da operadora, com planos pós ou pré-pago, desde que o aparelho seja compatível com a tecnologia Wap. O link da Gol está na seção Facilidades - Companhias Aéreas do Portal Tim Wap. Antes do primeiro acesso, é preciso cadastrar-se gratuitamente no site da Gol.

Pela navegação via celular, o cliente pagará de acordo com o plano contratado. A cobrança da passagem aérea é efetuada por uma das diversas formas de pagamento que o cliente pode escolher.

Mas uma pergunta que sempre recai sobre quem vai utilizar outro meio de mexer em dinheiro é a seguinte: estarei seguro? Estudo divulgado recentemente pela consultoria Tower Group mostrou que adeptos do sistema devem ficar atentos: vírus que atingem os chips dos telefones comprometem a segurança das transações bancárias.

Já foi identificado mais de 200 vírus de telefonia móvel, um número que duplica a cada seis meses. Para garantir a segurança das transações, a Tim e a Gol definiram que o login e a senha cadastrados no site serão solicitados todas as vezes que o acesso for feito pelo celular.

## 5. GRANDE CRESCIMENTO PARA O COMÉRCIO MÓVEL

por Marin Perez | InformationWeek EUA - 18/03/2009

Adoção de smartphones e adaptação de lojas como Amazon e eBay às plataformas móveis impulsionam modalidade

É crescente o número de pessoas que estão utilizando os celulares para transações financeiras e esses usuários devem gastar, em 2009, US$ 1,6 bilhão por meio do comércio móvel. A previsão é da ABI Research.

A pesquisa, chamada "Mobile Money Services And Contactless Payment Forecasts", revela que o crescimento é impulsionado por SMS, internet móvel e aplicativos para pagamento móveis. "Compras por meio da internet móvel está no centro das atenções", avaliou Mark Beccue, analista da empresa de pesquisa. "Com a adoção dos smartphones, um número grande de usuários estão comprando através dos portais mobile de empresas como Amazon e eBay", avisou.

A demanda por serviços de pagamento móveis cresce em todo o mundo. Em países industrializados, os usuários estão se familiarizando e se sentem confortáveis com a utilização dos celulares para mais funções além da voz. Em nações em desenvolvimento, os serviços financeiros móveis são encarados como forma de fugir dos bancos.

Esse movimento deve auxiliar empresas como PayPal, Obopay, Visa e Mastercard, que já deram os primeiros passos para ingressar no mercado móvel.

## 6. BANDA LARGA MÓVEL NO BRASIL

Apesar do grande crescimento por parte de fabricantes de dispositivos móveis e a oferta de serviços e produtos via web, as operadoras insistem em manter a banda larga em canais estreitos. Enquanto no Japão já se pode navegar a 42 Mbps no Brasil ainda esbarra-se na falta de infraestrutura e nas políticas nada convencionais fazendo com que as operadoras tenham um domínio quase que feudal sobre aqueles que utilizam banda larga.

> A Internet brasileira continua crescendo a passos largos. Os resultados apresentados pelo Centro de Estudos sobre as Tecnologias da Informação e da Comunicação - CETIC.br demonstram o avanço do acesso à grande rede, sobretudo por intermédio da banda larga. Os usuários brasileiros, cada vez mais numerosos, encontram na Internet uma fonte inesgotável de informações e conteúdos, bem como um ambiente de convívio através das redes sociais. No mundo observamos que os países buscam fortemente desenvolver políticas de massificação da infraestrutura de acesso à rede em alta velocidade. E a expansão do acesso via rede móvel é ponto central na soma de esforços destas políticas. O consenso é de que é preciso estar preparado para atuar com desenvoltura numa nova economia digital que cada vez mais é realidade imediata. (BECHARA, Marcelo. http://www.marcelobechara.com.br/home/?p=776 - 05/04/2010).

Segundo estudo, a banda larga móvel na América Latina soma 19 milhões de conexões. As operadoras sul-americanas mudaram seu foco, de serviços de voz (2G) para serviços de dados (3G), que oferecem serviços de valor agregado e alcançam usuários que consomem mais.

> "O uso da banda larga móvel contribuiu com o aumento do uso geral da banda larga na América Latina", disse Erasmo Rojas, diretor da 3G Américas para a América Latina e o Caribe. "A crescente participação de mercado e os 20 milhões de novas conexões no primeiro trimestre de 2010 indicam a oportunidade de crescimento em uma região que já representa 10% das assinaturas mundiais". (ROJAS, Erasmo - http://g1.globo.com/tecnologia/noticia/2010/05/banda-larga-movel-na-america-latina-e-caribe-soma-19-milhoes-de-conexoes.html)

Com base na demanda pela utilização da internet móvel, operadoras nacionais tem voltado seus olhos para estes clientes, oferecendo banda larga móvel a preços menores do que os praticados anteriormente, quando a banda larga móvel ainda era considerada recurso para classe alta e executivos . A operadora TIM, por exemplo, recentemente reduziu em cerca de 40% o valor do pacote de dados para celular. O novo Liberty Web, o assinante de plano pós-pago da TIM pode navegar pelo celular de forma ilimitada, enviar e receber e-mail, acessar redes sociais e se comunicar via aplicativos de mensagem instantânea por apenas R$ 29,90 ao mês. Existem oferta especiais para aqueles que compram um smartphone nas lojas TIM podendo receber até seis meses de internet ilimitada gratuita no celular. (http://www.grupocdn.com.br/lang/pt/sala-imprensa/tim-reduz-preco-de-acesso-a-internet-movel/)

Segundo a consultoria TELECO juntamente com a fabricante chinesa HUAWEI um levantamento realizado em junho último, nós três primeiros meses de 2010 o número de acessos em alta velocidade à internet por dispositivos sem fio no Brasil superou o volume de acessos fixos pela primeira vez. Foram 11,9 milhões de acessos móveis contra 11,8 milhões de acessos fixos.

> Do total de acessos móveis no primeiro trimestre, 8,7 milhões foram feitos por celulares enquanto o restante por modems. Apesar do crescimento vigoroso, "a densidade de banda larga no Brasil está abaixo da média mundial", afirma o levantamento, citando ainda que os preços da banda larga móvel no Brasil são maiores que os praticados na América Latina e Europa, "influenciados pela carga tributária e pelo subdimensionamento das redes, em especial em relação à capacidade das redes de transmissão". (REUTERS – 18/06/2010).

Já no 2º trimestre deste ano, os acessos de banda larga móvel somaram 13,9 milhões, divididos entre aparelhos celulares (10, 4 milhões) e modems (3,5 milhões). O crescimento da banda larga móvel foi de 17%, enquanto o da rede fixa atingiu 3% de crescimento. Ainda assim com o crescimento no período a densidade da banda larga móvel no Brasil ainda é pequena em relação ao resto do mundo.
De acordo com o estudo, no Brasil existem 3,6 conexões móveis para cada 100 habitantes, enquanto em todo o mundo há 9,5 conexões para 100 habitantes. A banda fixa possui 6 conexões para cada 100 habitantes no país, pouco abaixo da média mundial, que corresponde a 7,1 conexões para 100 habitantes.
Mesmo com o crescimento, 86% dos municípios (4.824 cidades) do país ainda não são atendidos pelo sinal 3G. (TABORDA, Cauã – INFO Online – Terça-feira, 24 de agosto de

2010 - http://info.abril.com.br/noticias/mercado/banda-larga-movel-cresce-34-1-em-12-meses-24082010-20.shl).

A internet móvel no Brasil ainda pode crescer mais, porém a alta carga tributária em serviços e equipamentos tem agido como freio reduzindo consideravelmente a velocidade de crescimento desse mercado no país. Contudo, diante de uma demanda cada vez maior por esse tipo de serviço, reduções mesmo que pequenas nos encargos recolhidos pelos fabricantes de celulares ao Fisco representariam um aumento surpreendente no número de assinantes de usuários da banda larga de terceira geração (3G), segundo a consultoria inglesa Telecom Advisory Services (TAS) e da GSMA, uma associação de indústrias do setor. Para eles, em um intervalo de cinco anos, a cada ponto percentual de corte nos impostos, 1 milhão de clientes a mais passariam a consumir o serviço. Do total pago por usuário da telefonia móvel brasileira, 43,3% são referentes a taxas e impostos. A redução desse valor para 42,3%, faria com que os ganhos para o setor e para a economia fossem significativos. Com mais usuários acessando a internet rápida, o governo também iria se beneficiar. No período de cinco anos, a receita com recolhimentos poderia crescer R$ 2,28 bilhões e o Produto Interno Bruto (PIB, soma de todas as riquezas do país) também ganharia, no mesmo prazo, um importante incremento de R$ 6,1 bilhões. (BRAGA, Fernando – CorreioBrasiliense http://www.correiobraziliense.com.br/app/noticia182/2010/07/24/economia,i=204161

## CONCLUSÃO

Como foi relatada neste artigo, a tendência é aumentar cada vez mais a aceitação da Tecnologia Móvel, como acontece no mundo inteiro. Como todos se rendem às tecnologias e principalmente ao comodismo, será muito rápida a implantação geral desse serviço bem como sua absorção. O crescimento é exponencial por questões óbvias e concretas, como por exemplo, o fato de não portar dinheiro em locais de alta ocorrência de roubos, o fato de conseguir encontrar pessoas que possam ser sequestradas, fácil localização de locais e estradas, receber notícias onde estiver e até mesmo fazer compras para casa ou trabalho sem ao menos ter que parar o que se está fazendo no momento, entre muitas outras coisas.

Percebe-se que a tecnologia móvel tem se tornado uma realidade no Brasil, assim como já é em países de primeiro mundo, devido à necessidade do ser humano em, através do uso da mobilidade, ter acesso à informações e serviços que facilitem o seu cotidiano.

**REFERÊNCIAS BIBLIOGRÁFICAS**

CASTELO, Marcelo, Tempo de Alerta, http://imasters.uol.com.br/artigo/14570

InfoMoney - 2007

LARA, Guilherme, Mobile Marketing terá crescimento vertiginoso em 2009, http://www.tramaweb.com.br/cliente_ver.aspx?ClienteID=204&NoticiaID=6316

PEREZ, Marin, Grande Crescimento para o Comércio Móvel, InformationWeek EUA – 2009.

Praestro Convergence - ebook Mobile Marketing: Mobile Marketing: Conceitos, Tecnologias e Cases, 2009.

BECHARA, Marcelo. http://www.marcelobechara.com.br/home/?p=776 – 05/04/2010

TABORDA, Cauã. http://info.abril.com.br/noticias/mercado/banda-larga-movel-cresce-34-1-em-12-meses-24082010-20.shl

REUTERS – 18/06/2010

BRAGA, Fernando – Correio Brasiliense
http://www.correiobraziliense.com.br/app/noticia182/2010/07/24/economia,i=204161